\documentclass[iop]{emulateapj}
\usepackage{epsfig}
\usepackage{apjfonts}
\usepackage{aas_macros}
\usepackage{color}

\begin{document}

\title{Inverse compton scattered merger-nova: late X-ray counterpart of gravitational wave signals from NS-NS/BH mergers}
\author{Shunke Ai$^{1}$ and He Gao$^{1,*}$}
\affiliation{
$^1$Department of Astronomy, Beijing Normal University, Beijing 100875, China; gaohe@bnu.edu.cn\\
}

\begin{abstract}

The recent observations of GW170817 and its electromagnetic (EM) counterparts show that double neutron star mergers could lead to rich and bright EM emissions. Recent numerical simulations suggest that neutron star and neutron star/black hole (NS-NS/BH) mergers would leave behind a central remnant surrounded by a mildly isotropic ejecta. The central remnant could launch a collimated jet and when the jet propagating through the ejecta, a mildly relativistic cocoon would be formed and the interaction between the cocoon and the ambient medium would accelerate electrons via external shock in a wide angle. So that the merger-nova photons (i.e., thermal emission from the ejecta) would be scattered into higher frequency via inverse compton (IC) process when they propagating through the cocoon shocked region. We find that the IC scattered component peaks at X-ray band and it will reach its peak luminosity in order of days (simultaneously with the merger-nova emission). With current X-ray detectors, such a late X-ray component could be detected out to 200 Mpc, depending on the merger remnant properties. It could serve as an important electromagnetic counterpart of gravitational wave signals from NS-NS/BH mergers. Nevertheless, simultaneous detection of such a late X-ray signal and the merger-nova signal could shed light on the cocoon properties and the concrete structure of the jet. 
\end{abstract}


\section {Introduction}

LIGO-VIRGO cooperation has reported five gravitational wave (GW) detection (GW150914, GW151226, GW170104, GW170608 and GW170814) from black hole-black hole (BH-BH) mergers, and one GW detection (GW170817) from neutron star-neutron star (NS-NS) merger, which means a new era of GW astronomy is coming \citep{abbott16a,abbott16b,abbott17,abbott17b,abbott17c,abbott17d}. Nevertheless, the electromagnetic (EM) counterparts of GW170817 have also been detected, including a weak short-duration gamma-ray burst (SGRB) \citep{goldstein17}, an optical/IR transient in the galaxy NGC 4993 \citep{coulter17}, a radio counterpart \citep{hallinan17} and a late re-brightening X-ray signal\citep{troja17}. A large number of teams across the world contributed to the observation of this source using ground- and space-based telescopes \cite[][for details]{abbott17e}.

Before the detection of GW170817, many associated EM counterparts have already been proposed for NS-NS/BH mergers, and their relative brightness is essentially determined by the properties of the post-merger central remnant object. In general, the merger remnant for NS-BH merger would be a BH. But NS-NS merger could lead to a BH or a supra-massive NS, depending on the total mass of the NS-NS system and the NS equation of state \citep{dai06,gaofan06,zhang13,lasky14,gao16}. It is generally believed that, for both cases, short duration gamma-ray bursts (sGRBs) and their afterglow emission are expected as one of the major EM counterparts of NS-NS/BH mergers \citep{eichler89,narayan92,berger14}. On the other hand, for both cases, the merger remnant would be surrounded by a mildly isotropic, sub-relativistic ejecta (which is composed of the tidally ripped and dynamically launched materials during the merger and the matter launched from the neutrino-driven wind from the accretion disk or neutron star surface \citep{rezzolla11,rosswog13,bauswein13,hotokezaka13,lei13,fernandez15,song17}). These ejecta are mostly composed of neutron-rich materials, and the radioactivity of these materials and the decay of r-process nuclei would heat the ejecta and then power an optical/IR transient \citep{li98,metzger10}.  When the merger remnant being a BH, r-process related radioactivity would be the only heating source, so that the luminosity of the optical/IR transient would be roughly $\sim10^3$ times of the nova luminosity \citep{metzger10}. But if the merger remnant being a supra-massive NS, its magnetic dipole radiation could serve as an additional heating sources to the ejecta \citep{zhang13,gao13}, which could easily exceed the r-process power, so that the thermal emission from the ejecta would be significantly enhanced \citep{yu13,metzgerpiro14}. The luminosity of the optical transient, in this case, would be systematically brighter by more than one order of magnitude than the r-process dominating cases \citep{gao17}. Since the thermal emission from the ejecta are essentially isotropic and also non-relativistic or mildly relativistic (due to the heavy mass loading) and therefore can be detected from any direction if the flux is high enough \cite[][for a review]{metzger17}.

All observed GW170817 EM counterparts are predicted, but some of them show unexpected behavior \citep[][for a review]{metzger17b}. For instance, the fluence ($\sim 1.4 \times 10^{-7} \ {\rm erg \ cm^{-2}} $) and spectral peak energy ($\sim 215$ keV) of the associated SGRB 170817A fall into the lower portion of the distributions of known sGRBs, but its peak isotropic luminosity ($\sim 1.7\times 10^{47}~{\rm erg \ s^{-1}}$) is abnormally low comparing with other SGRBs\citep{goldstein17}. Considering the relatively large upper limit of binary inclination angle relative to our line of site from the GW signal analysis \citep{abbott17d} and the self-consistency between $\gamma$-ray, X-ray, optical/IR and radio observations, a cocoon emission has been proposed as one of the most concordant model \citep{kasliwal17,piro17,gottlieb17b,xiao17}. Specifically, when a relativistic jet propagating through the surrounded ejecta, a mildly relativistic cocoon would be formed embracing the main jet, so that some relatively weak emission would be expected from a widen-angle structure located in the peripheral of the jet to explain the observed $\gamma$-ray emission \citep{gottlieb17,kathirgammaraju17,lazzati17a,lazzati17b}. Within this scenario, the X-ray and radio observations could be interpreted as the off-axis afterglow from the main jet or from the on-axis afterglow emission of the cocoon, depending on the concrete properties of the jet and the ejecta \citep{kasliwal17,piro17,gottlieb17b,troja17,guidorzi17}. 

If this interpretation is correct, probably all NS-NS mergers would generate a widen-angle mildly relativistic cocoon. The interaction between the relativistic cocoon and the ambient medium could generate an external shock, where particles are believed to be accelerated, giving rise to broad-band synchrotron radiation (cocoon afterglow emission). Considering the large opening angle of the cocoon, photons from other isotropic emission component, such as the thermal component from the ejecta, would be scattered into higher frequency via Inverse Compton (IC) process when they propagating through the cocoon external shock region. In this work, we will estimate the peak frequency and peak flux for this new emission component, and show the dependability of such emission by the currently available X-ray telescopes.


\section{Seed photons from ejecta thermal emission}

\subsection{Numerical model}

Considering that one NS-NS merger event leaves behind a central remnant (either BH or NS), surrounded by a neutron rich ejecta with mass $M_{\rm ej}$ and initial dimensionless speed $\beta_{\rm i}$. In any case, the ejecta would receive heating from the radioactive decay of the heavy nuclei synthesized in the ejecta via the r-process. If the central remnant is a NS, its magnetic dipole radiation could serve as an additional heating source to the ejecta. Moreover, the NS wind would continuously push from behind and accelerate the ejecta. Considering the energy dissipation through sweeping up the ambient medium, the dynamical evolution of the ejecta can be determined by \citep{yu13}
\begin{eqnarray}
{d\Gamma\over dt}={{dE\over dt}-\Gamma {\cal D}\left({dE'_{\rm int}\over
dt'}\right)-(\Gamma^2-1)c^2\left({dM_{\rm sw}\over dt}\right)\over
M_{\rm ej}c^2+E'_{\rm int}+2\Gamma M_{\rm sw}c^2},
\label{eq:Gt}
\end{eqnarray}
where $\Gamma$ is the bulk Lorentz factor, $t$ is time in the observer frame, ${\cal D}=1/[\Gamma(1-\beta)]$ is the Doppler factor, $E'_{\rm int}$ is the internal energy in the comoving frame, $t'$ is time in the comoving frame and $M_{\rm sw}=\frac{4\pi}{3}R^3nm_p$ is the shock swept mass from the interstellar medium (with density $n$), where $R$ is the radius of the ejecta in the lab frame.

With energy conservation, we have
\begin{eqnarray}
{dE\over dt}=L_{\rm cen}+{\cal D}^{2}L'_{\rm ra}-{\cal D}^{2}L'_{\rm e},
\label{eq:Et}
\end{eqnarray}
where $L_{\rm cen}$ is the injected energy from the central engine, $L'_{\rm ra}$ is the comoving radioactive power and $L'_e$ is the comoving radiated bolometric luminosity. When the central engine is a BH, we normally expect $L_{\rm cen}=0$ \cite[][for other opinion]{ma17}. When the central engine is a NS, normally a fraction $\xi$ of the dipole radiation luminosity is assumed to be injected into the ejecta \cite[][for details]{yu13}, i,e., $L_{\rm cen}=\xi L_{\rm d}$, where $L_d=L_{\rm sd}\left(1+{t\over t_{\rm sd}}\right)^{-2}$, with $L_{\rm sd}=10^{47}~R_{s,6}^6B_{14}^{2}P_{i,-3}^{-4}\rm~erg~s^{-1}$ being the spin down luminosity and $t_{\rm sd}=2\times10^{5}~R_{s,6}^{-6}B_{14}^{-2}P_{i,-3}^{2}~\rm s$ being the spin down timescale, where $P_{i}$, $B$ and $R_{s}$ are the initial spin period, the dipole magnetic strength and the radius of the NS. Throughout the paper, the convention $Q=10^n Q_n$ is used for cgs units, except for the ejecta mass $M_{\rm ej}$, which is in unit of solar mass $M_{\odot}$.

ere we adopt the empirical expression for the comoving radioactive power proposed by \cite{korobkin12}
\begin{eqnarray}
L'_{\rm ra}=4\times10^{49}M_{\rm ej,-2}\left[{1\over2}-{1\over\pi}\arctan \left({t'-t'_0\over
t'_\sigma}\right)\right]^{1.3}~\rm erg~s^{-1},
\label{eq:Lrap}
\end{eqnarray}
where $t'_0 \sim 1.3$ s and $t'_\sigma \sim 0.11$ s. The radiated bolometric luminosity could be expressed as
\begin{eqnarray}
L'_e=\left\{
\begin{array}{l l}
  {E'_{\rm int}c\over \tau R/\Gamma}, & \tau>1, \\
  {E'_{\rm int}c\over R/\Gamma}, &\tau<1,\\ \end{array} \right.\
  \label{eq:Lep}
\end{eqnarray}
where $\tau=\kappa (M_{\rm ej}/V')(R/\Gamma)$ is the optical depth of the ejecta with $\kappa$ being the opacity \citep{kasen10,kotera13}.

The variation of the comoving internal energy $E'_{\rm int}$ could be expressed as \cite[e.g.][]{kasen10,yu13}
\begin{eqnarray}
{dE'_{\rm int}\over dt'}=\xi {\cal D}^{-2}L_{\rm d}+ L'_{\rm ra} -L'_{\rm e}
-\mathcal P'{dV'\over dt'},
\label{eq:Ep}
\end{eqnarray}
where $\mathcal P'=E'_{\rm int}/3V'$ is the radiation dominated pressure. The comoving volume evolution can be fully addressed by $dV'/dt'=4\pi R^2\beta c$ together with $dR/dt=\beta c/ (1-\beta)$. 

With $E'_{\rm int}$ being solved, one can easily estimate the evolution of the effective temperature of the ejecta in the comoving frame $T'_{\rm eff}=(E'_{\rm int}/aV'\max(\tau,1))^{1/4}$. If a blackbody spectrum is assumed for the thermal emission from the ejecta, one can calculate the observed flux for a given frequency $\nu$
\begin{eqnarray}
F_{\nu}={1\over4\pi D_L^2}{8\pi^2  {\cal D}^2R^2\over
h^3c^2\nu}{(h\nu/{\cal D})^4\over \exp(h\nu/{\cal D}kT'_{\rm eff})-1},
\end{eqnarray}
where $k$ is the Boltzmann constant, $a$ is the radiation constant, $h$ is the Planck constant and $D_L$ is the luminosity distance.

\subsection{Analytical estimation}
In principle, one can apply above numerical modeling to calculate the observed flux of the thermal emission from the ejecta (henceforth we call it merger-nova emission) at any time and any frequency. The merger-nova photons could serve as the seed photons to be scattered by the cocoon-medium shock into X-ray band, which will be discussed later in detail. In order to better understand the features of these seed photons, we present some analytic approximation for the results of the merger-nova emission, such as the peaking time of the merger-nova, the peak frequency and peak flux at that time.

\subsubsection{Black hole as merger remnant}
When the merger remnant is a BH, following the analytical estimation from \cite{metzger10}, the bolometric luminosity of the r-process-powered merger-nova would reach its peak when the photon diffusion time-scale equals the expansion time-scale, where the radius of ejecta being as    
\begin{eqnarray}
     R_{p}&=&\left(\frac{Bv_{\rm ej}{\kappa}M_{\rm ej}}{c}\right)^{1/2}\nonumber
     \\
     &\approx & 3.74\times10^{14}{\rm cm}~\left(\frac{v_{\rm ej}}{0.1c}\right)^{1/2} \left(\frac{M_{\rm{ej}}}{10^{-2}M_{\odot}}\right)^{1/2}\left(\frac{\kappa}{1~{\rm cm^2~g^{-1}}}\right)^{1/2},
    \end{eqnarray}
where $M_{\rm ej}$, $v_{\rm ej}$ and $\kappa$ are the mass, velocity and opacity of the ejecta, and $B\approx 0.07$ for spherical outflow \cite[e.g.,][]{padmanabhan00}. Under the assumption of free expansion, $R_{p}$ would be reached on a time-scale
    \begin{equation}
    t_{p}=1.25\times 10^{5}{\rm s}~\left(\frac{v_{\rm ej}}{0.1c}\right)^{-1/2}\left(\frac{M_{\rm{ej}}}{10^{-2}M_{\odot}}\right)^{1/2}\left(\frac{\kappa}{1~{\rm cm^2~g^{-1}}}\right)^{1/2}.
    \end{equation}
The peak bolometric luminosity is given as
    \begin{eqnarray}
    \nonumber
    L_{p}=&1.44&\times10^{41}{\rm erg~s^{-1}}\left(\frac{f}{10^{-6}}\right)\left(\frac{v_{\rm ej}}{0.1c}\right)^{1/2}
    \\
    &\times &\left(\frac{M_{\rm{ej}}}{10^{-2}M_{\odot}}\right)^{1/2}\left(\frac{\kappa}{1~{\rm cm^2~g^{-1}}}\right)^{-1/2},
    \end{eqnarray}
where $f$ is a dimensionless number characterizing the heating generation rate \citep{li98,metzger10}.

Using Stefan-Boltzmann law, the effective temperature is roughly estimated as
    \begin{eqnarray}
    \nonumber
    T_{p}=&6.17&\times10^{3}{\rm K}~\left(\frac{f}{10^{-6}}\right)^{1/4}\left(\frac{v_{\rm ej}}{0.1c}\right)^{-1/8}
    \\
    &\times &\left(\frac{M_{\rm{ej}}}{10^{-2}M_{\odot}}\right)^{-1/8}\left(\frac{\kappa}{1~{\rm cm^2~g^{-1}}}\right)^{-3/8}.
    \end{eqnarray}
Assuming a blackbody spectrum, we can estimate the peak frequency of the thermal emission as
    \begin{eqnarray}
    \nu_{p}= &3.63&\times10^{14}{\rm Hz}~\left(\frac{f}{10^{-6}}\right)^{1/4}\left(\frac{v_{\rm ej}}{0.1c}\right)^{-1/8}\nonumber
    \\
    & \times & \left(\frac{M_{\rm{ej}}}{10^{-2}M_{\odot}}\right)^{-1/8}\left(\frac{\kappa}{1~{\rm cm^2~g^{-1}}}\right)^{-3/8}.
    \end{eqnarray}
The according peak flux could be given as
    \begin{eqnarray}
    F_{\nu_{p}}=&3.15&\times10^{2}{\rm \mu Jy}~\left(\frac{f}{10^{-6}}\right)^{3/4}\left(\frac{v_{\rm ej}}{0.1c}\right)^{5/8}\nonumber
\\
    &\times & \left(\frac{M_{ej}}{10^{-2}M_{\odot}}\right)^{5/8}\left(\frac{d}{10^{26}}\right)^{-2}\left(\frac{\kappa}{1~{\rm cm^2~g^{-1}}}\right)^{-1/8}.
    \end{eqnarray}
    \subsubsection{Massive neutron star as merger remnant}
    
If the equation of state of nuclear matter is stiff enough, the central product for a binary neutron star merger could be a stable or a supra-massive NS rather than a black hole. This newborn massive NS would be rotating with a rotation period on the order of milliseconds, and may also contain a strong magnetic field $B\gtrsim10^{14}$ G similar to ``magnetars". In this case, the magnetar dipole radiation could easily dominate the heating and accelerating process for the ejecta, so that the dynamics of the ejecta could be defined by energy conservation \cite[][for details]{gao13}
\begin{eqnarray}
\xi L_d t=(\Gamma-1)M_{\rm ej}c^2+(\Gamma^2-1)M_{\rm sw}c^2. 
\end{eqnarray}
Considering that the number density of ambient medium in the NS-NS merger scenario should be usually low, for most situations, the deceleration time for the ejecta should be larger than the spin down timescale of the NS. At this stage, we have $(\Gamma-1)M_{\rm ej}c^2>>(\Gamma^2-1)M_{\rm sw}c^2$, so that $\Gamma-1 \propto t$. With the energy injection from the millisecond magnetar, the ejecta could be accelerated into a mildly relativistic or even relativistic speed \citep{gao13}. In this case, we have  the approximation of $\Gamma \propto t$. With $dR/dt=(\beta c)/(1-\beta)$, we have $R\propto t^{3}$.  

The merger-nova emission peaks at $t_p$ when $\tau\sim 1$, where $\tau=\kappa\left(M_{\rm{ej}}/V\right)R$ is defined as the optical depth of the ejecta, and $\Gamma$ is the bulk Lorentz factor of the ejecta, $R$ and $V$ are the radius and volume of the ejecta in the lab frame. The radius of the ejecta at $t_p$ could be simply estimated as

\begin{equation}
R_{p}\sim 6.92\times 10^{14}{\rm cm}~\kappa^{1/2}M_{\rm{ej},-3}^{1/2}.
\end{equation}

When $t_{\rm{sd}}>t_{p}$, we can estimate the peak time in the observer frame as
\begin{eqnarray}
t_{p}&=&\left({R_p\over R_{\rm sd}}\right)^{1/3}t_{\rm sd}\nonumber
\\
&\sim& 1.54\times 10^{4}{\rm s}~\kappa^{1/6}\xi^{-2/3}M_{\rm{ej},-3}^{5/6}B_{p,14}^{-4/3}P_{0,-3}^{8/3}R_{s,6}^{-4},
\end{eqnarray}
where $R_{\rm sd}\sim 2\Gamma^2_{\rm sd}ct_{\rm sd}$ is the radius of the ejecta at spin-down timescale and $\Gamma_{sd}\sim \xi E_{\rm rot}/M_{\rm ej}c^2$ is the Lorenzt factor of the ejecta at the spin-down timescale and $E_{\rm{rot}}=(1/2)I \Omega_{0}^{2} \simeq 2\times 10^{52} I_{45} P_{0,-3}^{-2} ~{\rm erg}$ (with $I_{45} \sim 1.5$ for a massive neutron star) is the total spin energy of the millisecond magnetar.

At $t_{p}$, the effective temperature of the ejecta in the observer frame would be
\begin{eqnarray}
T_{p}&=&({\xi L_d\over 4\pi R^2_p \sigma})^{1/4}\nonumber
\\
&\sim& 1.31\times 10^{5}{\rm K}~\kappa^{-1/4}\xi^{1/4}M_{\rm{ej},-3}^{-1/4}B_{p,14}^{1/2}P_{0,-3}^{-1}R_{s,6}^{3/2},
\end{eqnarray}
where we assume a constant fraction ($\xi$) of the NS dipole luminosity would be thermalized in the outflow. With the numerical modeling, it is found that around the transparent time (i.e., $\tau=1$), the radiated bolometric luminosity of the merger-nova is close to the injection energy power $\xi L_{d}$ \citep{yu13}.

Assuming a blackbody spectrum, we can estimate the peak frequency of the thermal emission in the observer frame as
\begin{eqnarray}
\nu_{p} \sim  7.70\times 10^{15}{\rm Hz}~\kappa^{-1/4}\xi^{1/4}M_{\rm{ej},-3}^{-1/4}B_{p,14}^{1/2}P_{0,-3}^{-1}R_{s,6}^{3/2}.
\end{eqnarray}
The according peak flux could be given by
\begin{eqnarray}
   F_{\nu_{p}} &\sim&  1.03\times 10^7 {\rm \mu Jy}~\kappa^{1/4}\xi^{3/4}M_{\rm{ej},-3}^{1/4}B_{p,14}^{3/2}\nonumber
\\
&~&\times P_{0,-3}^{-3}R_{s,6}^{9/2}d_{26}^{-2}.
\end{eqnarray}

Similarly, for $t_{\rm{sd}}<t_{p}$ case, we have $\Gamma \propto t$ and $R \propto t^{3}$ when $t<t_{\rm{sd}}$ and $\Gamma\propto t^{0}$ and $ R\propto t$ when $t>t_{\rm{sd}}$ \citep{gao13}, where $L_{d}\approx L_{\rm{sd}}\left({t}/{t_{\rm{sd}}}\right)^{-2}$ is taken when $t>t_{\rm{sd}}$. Then we can estimate the peak time of mergernova as

\begin{eqnarray}
&t_{p}&\sim 4.07 \times 10^{3}{\rm s}~ \kappa^{1/2}\xi^{-2}M_{\rm{ej},-3}^{5/2}P_{0,-2.5}^{4}.
\end{eqnarray}

In this case, the temperature of the ejecta at $t_{p}$ would be
\begin{eqnarray}
T_{p}&\sim  2.00\times 10^{5}{\rm K}~\kappa^{-1/2} \xi^{5/4} M_{\rm{ej},-3}^{-3/2}B_{p,15.5}^{-1/2}P_{0,-2.5}^{-3}R_{s,6}^{-3/2}.
\end{eqnarray}

The peak frequency of the merger-nova emission in the observer frame is
\begin{eqnarray}
\nu_{p} \sim  1.18\times 10^{16}{\rm Hz}~\kappa^{-1/2}\xi^{5/4} M_{\rm{ej},-3}^{-3/2}B_{p,15.5}^{-1/2}
P_{0,-2.5}^{-2}R_{s,6}^{-3/2},
\end{eqnarray}
and the corresponding peak flux is
\begin{eqnarray}
F_{\nu_{p}} &\sim&  3.66\times 10^{7} {\rm \mu Jy}~\kappa^{-1/2}\xi^{15/4}M_{\rm{ej},-3}^{-7/2}B_{p,15.5}^{-3/2}\nonumber
\\
&~& \times P_{0,-2.5}^{-6}R_{s,6}^{-9/2}d_{26}^{-2}.
\end{eqnarray}

    \subsection{Inverse Compton scattering}

During the propagation of the cocoon, an external shock would form upon interaction with the ambient medium. The shock-accelerated electrons behind the blast wave are usually assumed to be distributed with a power-law function of electron energy, with a minimum Lorentz factor $\gamma_m$: $N(\gamma_{e})d\gamma_{e} \propto \gamma_{e}^{-p}d\gamma_{e},\gamma_{e}\geq\gamma_{m}$.
Assuming that a constant fraction $\epsilon_e$ of the shock energy is distributed to electrons, the
minimum injected electron Lorentz factor can be estimated as
\begin{equation}
    \gamma_{m}=\epsilon_{e} \left(\frac{p-2}{p-1}\right)\frac{m_{p}}{m_{e}} \Gamma_{\rm co}\cong42.2~\epsilon_{e,-2}\Gamma_{\rm co,1}, 
\end{equation} 
where $\Gamma_{\rm co}$ is the Lorentz factor of the cocoon, and $p=2.3$ is adopted as commonly used in GRB afterglow modeling \citep{kumarzhang15}.
    
When the seed photons (with frequency $\nu$) from the merger-nova propagate through the cocoon external shock region, they would be scattered into higher frequency via IC process. The typical photon frequency of the IC scattered merger-nova would be estimated as\footnote{In principle, one needs to firstly transform the seed photon energy from the observer frame to the cocoon-medium shock front frame, then calculate the inverse Compton scattering in that frame, and finally transform the scattered photon energy back to the observer frame. For simplicity, we assume that the seed photon injection direction is aline with the moving direction of the cocoon-medium shock front, in which case the two step relativistic transformations could be canceled out.} (Sari \& Esin 2001)

\begin{eqnarray}
\nu_p^{IC}=2\gamma^2_m v_p.
\end{eqnarray}

When merger remnant is a BH, we have 

\begin{eqnarray}
    \nu_{p,{\rm BH}}^{IC} &\sim & 1.29\times 10^{18}{\rm Hz}~\epsilon_{e,-2}^{2}\left(\frac{f}{10^{-6}}\right)^{1/4}\left(\frac{v_{\rm ej}}{0.1c}\right)^{-1/8}
    \nonumber
    \\
    &\quad &\times\left(\frac{M_{\rm ej}}{10^{-2}M_{\odot}}\right)^{-1/8}\left(\frac{\kappa}{1~{\rm cm^2~g^{-1}}}\right)^{-3/8}.
    \end{eqnarray}
When the merger remnant is a NS, we have
    \begin{eqnarray}
    \nu_{p,{\rm NS}}^{IC} &\sim & 2.74\times 10^{19}\rm{Hz}~\epsilon_{e,-2}^{2}\,\kappa^{-1/4}\xi^{1/4}
    \nonumber
\\
  &\quad &\times  M_{\rm{ej},-3}^{-1/4}B_{p,14}^{1/2}P_{0,-3}^{-1}R_{s,6}^{3/2},
    \end{eqnarray}    
when $t_{\rm{sd}}>t_{p}$ and    
\begin{eqnarray}   
    \nu_{p,{\rm NS}}^{IC} &\sim & 4.20\times 10^{19}\rm{Hz}~\epsilon_{e,-2}^{2}~\kappa^{-1/2}\xi^{5/4} M_{ej,-3}^{-3/2}\nonumber
\\
  &\quad & \times B_{p,15.5}^{-1/2}P_{0,-2.5}^{-2}R_{s,6}^{-3/2},
    \end{eqnarray}
when $t_{\rm{sd}}<t_{p}$.

The peak flux for the IC scattering component could be estimated as 
    \begin{equation}
    F_{\nu_{p}}^{IC}=\tau^{IC}F_{\nu_{p}}
    \end{equation}
where $\tau^{IC}=\left(\frac{1}{3}nR\sigma_{T}\right)$ is the optical depth for IC scatterings for a constant density medium \citep{sari01}, and $\sigma_{T}$ is Thompson scattering cross section. We thus have
    \begin{eqnarray}
    F_{\nu_{p},{\rm BH}}^{IC}&\sim & 2.62\times10^{-5} {\rm \mu Jy}~\left(\frac{f}{10^{-6}}\right)^{-3/4}\left(\frac{v_{\rm ej}}{0.1c}\right)^{1/8}\nonumber
\\    
    &\quad &\times\left(\frac{M_{\rm ej}}{10^{-2}M_{\odot}}\right)^{9/8} \left(\frac{d}{10^{26}}\right)^{-2}\nonumber
\\
&~&\times \left(\frac{\kappa}{1~{\rm cm^2~g^{-1}}}\right)^{3/8}\left(\frac{n}{1~{\rm cm}^{-3}}\right),
\end{eqnarray}
and 
\begin{eqnarray}
F_{\nu_{p},{\rm NS}}^{IC}&\sim & 2.11\times10^{-1} \rm{\mu Jy}~\kappa^{5/12}\xi^{1/12}M_{\rm{ej},-3}^{13/12}\nonumber
\\
&\quad &\times B_{p,14}^{1/6}P_{0,-3}^{-5/3}R_{s,6}^{1/2}nd_{26}^{-2},
\end{eqnarray}
and
\begin{eqnarray}
F_{\nu_{p},{\rm NS}}^{IC}&\sim & 1.98\times10^{-1} \rm{\mu Jy}~\xi^{7/4} M_{\rm{ej},-3}^{-1}\nonumber
\\
&\quad &\times B_{p,15.5}^{-3/2}P_{0,-2.5}^{-2}R_{s,6}^{-9/2}nd_{26}^{-2}
\end{eqnarray}
for $t_{\rm{sd}}>t_{p}$ case and $t_{\rm{sd}}<t_{p}$ case respectively.

\section{Detectabilty} \label{sec:floats}
We use the numerical method described in section 2.1 to calculate the light curve of the inverse compton scattered merger-nova emission. Given some fiducial parameters, such as the ejecta mass, velocity, opacity being as $10^{-2}M_{\odot}$, $0.2c$, and $1{\rm cm^{-2}g^{-1}}$, and the magnetar parameters being as $B=10^{14}\rm{G}$, $P_{0}=1\rm{ms}$, $R_s=10^6~{\rm cm}$ and $\xi=0.3$, we find that the IC scattered merger-nova peaks at X-ray band. In figure 1, we compared the peak flux of the IC scattered merger-nova with the sensitivity of current X-ray facilities, such as Swift/XRT, Chandra and XMM-Newton \citep{burrows05,weisskopf02,jasen01}. We find that  
for NS-BH mergers or NS-NS mergers with BH being the merger remnant, such a X-ray component is only detectable out to 2 Mpc with current available facilities. However, for NS-NS mergers with massive NS being the merger remnant, the IC scattered X-ray component would become much brighter, and it will be detectable out to 200 Mpc (the designed horizon of aLIGO for NS-NS mergers \citep{abbott09}). In figure 1, we also plot the expected IC scattered merger-nova for GW170817, and we find that it is too dim to account for the late Chandra X-ray observations.

\begin{figure}[ht!]
\plotone{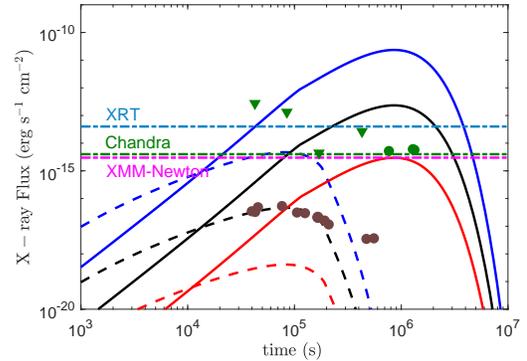}
\caption{The light curve of IC scattered merger-nova in X-ray. The soild lines represent the cases when seed photon is magnetar-powered merger-nova, while the dash lines represent the cases when seed photon is r-process-powered merger-nova. Different colors stand for different distance assumption (blue for $d\sim 2\rm{Mpc}$; black for $d\sim 20\rm{Mpc}$; red for $d\sim 200\rm{Mpc}$). Dotted dash lines are the sensitivities of currently available X-ray telescopes. Here the mass of the ejecta is taken as $10^{-2}M_{\odot}$. The opacity is taken as $\kappa=1 \rm{cm^{-2}\,g^{-1}}$. The initial velocity of ejecta is adopted as $\beta_{i}=0.2$. The magnetar parameters are taken as $B_{p}=10^{14}\rm{G}$, $P_{0}=1\rm{ms}$, $R_{s}=10^{6}\rm cm$ and $\xi=0.3$. The initial bulk Lorentz factor of the cocoon is assumed as $\Gamma_{\rm co}=10$. The green dots represent the observational data in X-ray for GW170817 \citep{troja17}, where the inverted triangles stand for observational upper limits. The brown dots represent the expected IC scattered merger-nova emission for GW170817. \label{fig:general}}
\end{figure}

\section{Conclusion and discussion} 

The recent observations of GW170817 and its EM counterparts have proven the prediction that NS-NS/BH mergers could lead to rich and bright EM emissions, invoking several emission components. In general, the jet component would give rise to a SGRB and its afterglow, and the isotropic component would give rise to a merger-nova emission and its afterglow. Recent studies suggest that the jet component is structured, with a relativistic jet surrounded by a mild relativistic cocoon. In this case, the interaction between the cocoon and the ambient medium would accelerate electrons via external shock in a wide angle. So that the merger-nova photons would be scattered into higher frequency via IC process when they propagating through the cocoon external shock region. 

In this work, we find that the IC scattered component peaks at X-ray band and it will reach its peak luminosity simultaneously with the merger-nova. For NS-BH mergers or NS-NS mergers with BH being the merger remnant, the X-ray component is detectable out to 2 Mpc with current facilities (such as Chandra and XMM-Newton). On the other hand, if the total mass of binary neutron star system is small enough and the equation of state of nuclear matter is stiff enough, the merger of two NSs could leave behind a supra-massive NS. In this case, the merger-nova emission could be significantly enhanced, so that the IC scattered X-ray component also becomes brighter. It will be detectable out to 200 Mpc with current facilities. Note that even for BH remnant case, the magnetic wind driven by Blandford-Payne process \citep{BP} from new-born BH accretion disk or fallback accretion disk would significantly enhance the merger-nova emission \cite[][for details]{chen17,ma17}, in this case, the IC scattered X-ray component could become as bright as the magnetar remnant case. 

Our newly proposed late X-ray emission could serve as an important EM counterpart of GW signals. Simultaneous detection of such X-ray signal and the merger-nova signal could help to investigate the cocoon properties and the concrete structure of the jet. 

It is worth noticing that some other mechanism could also generate late X-ray emission, which may outshine the proposed signal here. For instance, if the merger remnant of NS-NS is a supra-massive NS, the X-rays powered by NS wind dissipation would diffuse out at late time when the ejecta becomes (or be close to) optically thin, a late X-ray re-brightening would be expected \citep{metzgerpiro14,gao15,gao17}. But if the supra-massive NS have collapsed into a black hole before the surrounding ejecta becomes transparent, such signal would disappear. On the other hand, the afterglow emission from the structure jet could also provide X-ray photons, but its strength is sensitively depending on the viewing angle and the energy distribution within the jet. For most proper viewing angles, the corresponding jet energy is usually small, so that a relatively weak X-ray afterglow emission is expected \citep{lazzati17b}.

\acknowledgments

This work is supported by the National Basic Research Program (973 Program) of China (Grant No. 2014CB845800), the National Natural Science Foundation of China under Grant No. 11722324, 11603003, 11633001 and 11690024, and the Strategic Priority Research Program of the Chinese Academy of Sciences, Grant No. XDB23040100.

\end{document}